\documentclass[aps, prl, reprint, altaffilletter, floatfix]{revtex4-1}
\usepackage{graphicx}
\usepackage{float}
\usepackage{amsmath}
\usepackage{color}
\usepackage{multirow}

\begin{document}


\title{
Direct asymmetry measurement of temperature and density spatial distributions in inertial confinement fusion plasmas from pinhole space-resolved spectra
} 



\author{T. Nagayama}
\altaffiliation[Present address: ]{Sandia National Laboratories, Albuquerque, NM}
\affiliation{Physics Department, University of Nevada, Reno, NV}

\author{R. C. Mancini}
\affiliation{Physics Department, University of Nevada, Reno, NV}

\author{R. Florido}
\altaffiliation[Present address: ]{Departamento de F\'isica, Universidad de Las Palmas de Gran Canaria, 35017 Las Palmas de Gran Canaria, Spain}
\affiliation{Physics Department, University of Nevada, Reno, NV}

\author{D. Mayes}
\affiliation{Physics Department, University of Nevada, Reno, NV}

\author{R. Tommasini}
\author{J. A. Koch}
\affiliation{Lawrence Livermore National Laboratory, Livermore, CA}

\author{J. A. Delettrez}
\author{S. P. Regan}
\affiliation{Laboratory for Laser Energetics, University of Rochester, NY}
\author{V. A. Smalyuk}
\altaffiliation[Present address: ]{Lawrence Livermore National Laboratory, Livermore, CA}
\affiliation{Laboratory for Laser Energetics, University of Rochester, NY}


\date{\today}

\begin{abstract}
Two-dimensional space-resolved temperature and density images of an inertial confinement fusion (ICF) implosion core have been diagnosed for the first time. 
Argon-doped, direct-drive ICF experiments were performed at the Omega Laser Facility and a collection of two-dimensional space-resolved spectra were obtained from an array of gated, spectrally resolved pinhole images recorded by a multi-monochromatic x-ray imager.  
Detailed spectral analysis revealed asymmetries of the core not just in shape and size but in the temperature and density spatial distributions, thus characterizing the core with an unprecedented level of detail. 
\end{abstract}

\pacs{}

\maketitle 

Inertial confinement fusion (ICF) is an approach that utilizes laser
produced ablation pressure to compress a millimeter-sized spherical
shell capsule containing fuel (e.g., deuterium and tritium) and drive
the fuel temperature and density to conditions suitable for \textit{ignition}
\cite{Lindl:2004jz}. The key is a spherically symmetric and stable
compression. While state-of-the-art hydrodynamics simulations have
been used to design ignition implosions, the challenge of achieving
a symmetric implosion experimentally has thus far prevented ICF from
reaching the conditions required for successful ignition \cite{Landen:2012bq}.
Hence, measuring the spatial asymmetry in the temperature and density
distributions in the implosion core is crucial for understanding how
to make it more symmetric.

Several diagnostics were developed in the last few decades in order
to investigate implosion core conditions. K-shell line emission spectroscopy
using Ar as a tracer has proved to be a powerful tool to extract space-averaged
electron temperature, $T_{e}$, and density, $n_{e}$ \cite{Griem:1992jv,Hammel:dg,Bailey:2004bq}.
However, two-dimensional (2-D) space-resolved spectra have never been
extracted to study the asymmetries of $T_{e}$ and $n_{e}$ structures
in the implosion core. X-ray pinhole imaging of ICF implosion cores
has been used to study the shape and size of the core and, in particular,
to characterize deviations from spherical symmetry \cite{Suter:1994bz,Murphy:1998fk}.
Nevertheless, these images do not reveal the implosion asymmetries
in $T_{e}$ and $n_{e}$ distributions.

This Letter describes a new spectroscopic method that combines the
ideas of Ar tracer spectroscopy and pinhole imaging to extract implosion
core images in $T_{e}$ and $n_{e}$ without making symmetry assumptions.
These pinhole images are extracted by analyzing a collection of 2-D
space-resolved spectra obtained from an array of spectrally resolved
core images. The direct measurement of temperature and density asymmetries
in the core provides stringent constraints on what actually happens
in implosion experiments and can be used to benchmark hydrodynamic
simulations. The discussion here focuses on the application to ICF
implosion core conditions; nevertheless, the ideas are general. The
extraction and analysis of space-resolved spectra from spectrally
resolved pinhole images opens up new possibilities for x-ray spectroscopy
of high-energy density plasmas.

\begin{figure}
\includegraphics[width=8.5cm]{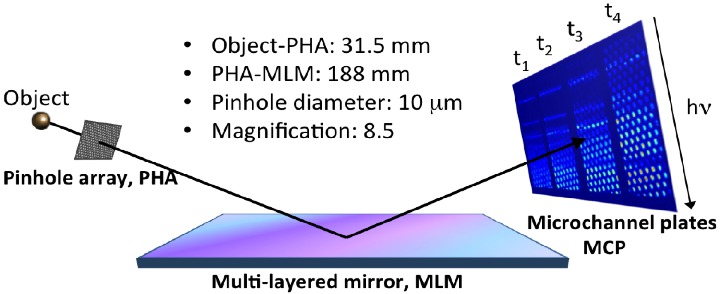}

\caption{\label{fig:MMI schematic} Schematic of the MMI, which consists of
a pinhole array, a multi-layered mirror, and microchannel plates (not
to scale).}
\end{figure}

The spectroscopic data were recorded in a series of Ar-doped ICF implosion
experiments performed at the Omega Laser Facility of the Laboratory
for Laser Energetics. In the experiments, 60 OMEGA laser beams ($\lambda\approx350\,\mathrm{nm}$,
total energy $\mathrm{\approx18\, kJ}$, pulse-duration $\approx$
2 ns) were used to irradiate the surface of spherical plastic shell
(radius $\mathrm{\approx400\,\mu m}$, shell thickness $\mathrm{\approx27\,\mu m}$)
filled with 20 atm of $\mathrm{D_{2}}$. LILAC 1-D \cite{Delettrez:1987fn}
hydrodynamic simulations predict that the implosion core plasma conditions
would range 1-3 keV in $T_{e}$ and $1-4\times10^{24}$ $\mathrm{cm^{-3}}$
in $n_{e}$ around the collapse of the implosion. Ar K-shell spectroscopy
is appropriate to diagnose these conditions \cite{Griem:1992jv,Hammel:dg,Bailey:2004bq}.
Spectral line broadening of the Ar $\mathrm{Ly\beta}$ and $\mathrm{He\beta}$
(i.e., n=3 to 1 transitions in H-like and He-like Ar, respectively)
is dominated by the Stark effect, which is sensitive to $n_{e}$.
The line intensity ratio of Ar $\mathrm{Ly\beta}$ to $\mathrm{He\beta}$
is dependent on $T_{e}$. Thus, a tracer amount of Ar (0.18 \% atomic
concentration) was mixed into the fuel to infer the implosion core
plasma conditions. Previous observations of space-integrated Ar K-shell
spectra have uniquely determined $T_{e}$ and $n_{e}$ of implosion
core plasmas by fitting the data with detailed spectral models \cite{Griem:1992jv,Hammel:dg,Bailey:2004bq}.
In order to investigate the asymmetry in $T_{e}$ and $n_{e}$ spatial
distributions, the challenge is to record Ar line emission not just
with spectral resolution, but also with 2-D spatial resolution.

The multi-monochromatic x-ray imager (MMI)\cite{2006SPIE.6317E..34T}
is a unique 2-D space-resolved x-ray spectrometer, which consists
of a pinhole array (PHA), a multi-layered Bragg mirror (MLM), and
microchannel plates (MCP) (Fig. 1). Characteristic parameters of the
instrument are listed in the figure. For the experiments discussed
here, the four MCP strips were triggered at different times with an
interstrip delay of 100 ps, thus recording four snapshots of spectrally
resolved 2-D implosion core image arrays ($\mathrm{\Delta x=10\,\mu m}$).
Figure 2(a) shows a blow up of one MCP strip ($t_{3}$ in Fig. 1).
Photons passing through different pinholes have slightly different
incident angles onto the mirror, which produces spectral resolving
power, $E/\Delta E$, of 150 in the horizontal direction of Fig. 2(a)
(i.e., vertical direction of the MCP strip in Fig. 1). On the MCP,
the $He\beta$ and $Ly\beta$ lines are approximately separated by
20 ps in time. More details about the instrument are described elsewhere
\cite{2006SPIE.6317E..34T}. These spectrally resolved images are
rich in information, and can be processed to produce broad- and narrow-band
images and the space-integrated spectrum \cite{Welser:2003fm}.

\begin{figure}
\includegraphics[width=8.5cm]{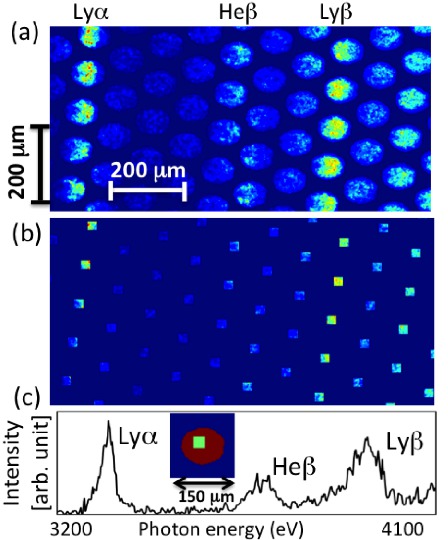} \caption{\label{fig:MMI data}(a) MMI data from the $t_{3}$ MCP strip in Fig.
\ref{fig:MMI schematic}, which records spectrally resolved pinhole
images of the implosion core. In addition to the 2-D spatial resolution,
the horizontal axis is also a spectral axis. (b) MMI data associated
with a sub-region (green square) created by applying a core sub-region
mask to the original MMI data {[}i.e., (a){]}. (c) space-resolved
spectrum for the sub-region extracted from (b).}
\end{figure}

Recently, there have been major improvements in MMI data processing,
including a more accurate and efficient technique for the determination
of image centers \cite{Nagayama:2011il}. The artifacts associated
with the discrete nature of the data were significantly suppressed
both on the space-integrated spectrum and quasi-monochromatic images.
An implosion core mask was introduced to limit the background signal
and to mitigate the spatial bias in the sampled points at each photon
energy. This dramatically improved the quality of the space-integrated
spectrum. In this Letter, we demonstrate for the first time that 2-D
space-resolved spectra can be extracted from the spectrally resolved
images. First, we define a mask associated to a sub-region of the
core image whose location is specified relative to image center {[}e.g.,
the green square of Fig. 2(c){]}. Then, we apply the sub-region mask
to each individual core image on the original MMI data and pick out
the data associated with the selected sub-region {[}Fig. 2(b){]}.
Finally, we compute the spectrum associated with this sub-region using
the technique discussed in Ref. \cite{Nagayama:2011il}. The resultant
spectrum shown in Fig. 2(c) is a 2-D space-resolved spectrum formed
by photons only from the selected sub-region of the core image. The
size of the spatial region is limited by the spatial resolution of
the instrument and signal-to-noise ratio (S/N).

By repeating the same procedure for different sub-regions, one can
extract a collection of space-resolved spectra, which are formed by
radiation emitted from different sub-volumes (or \textit{chords} of
finite cross-section) in the implosion core. Figure 3(a) is a schematic
to visualize the connection between a sub-region on the image plane
and the corresponding sub-volume in the implosion core that produces
the space-resolved spectrum extracted for the selected sub-region.
The chord associated with a space-resolved spectrum can be defined
by the intersection of the core volume (i.e., a volume indicated by
a dashed line) and the projection of the spatial region along the
line of sight {[}Fig. 3(a){]}. This parallel ray tracing assumption
is a good approximation. The rays collected by a single pinhole are
parallel within arctan($\mathrm{D_{PH}}$/$D_{0}$)$<$0.02$^{\circ}$
where $D_{PH}$=10 $\mu$m is the pinhole diameter and $D_{0}$=$3.15$
cm is the object to pinhole distance. Signals through different pinholes
deviate more from parallel. However, over the He$\beta$ and Ly$\beta$
regions of the MMI data, the pinholes are separated by less than $400\,\mu\mathrm{m}$
(the separation between adjacent pinholes is 90 $\mu$m). The rays
going through pinholes separated by 400 $\mu$m deviate from parallel
by less than $1^{\circ}$ {[}i.e., $\sim$arctan(400 $\mu$m/$D_{0}$){]}.
At maximum chord length (i.e., $\sim$100 $\mu$m), this deviation
results in an error on the magnification-corrected image plane of
$\mathrm{100\,\mu m}\times\tan(1^{\circ})=1.7\,\mu m$, which is much
smaller than the spatial resolution of the instrument and thus negligible.

The photon mean-free-path of the diagnostic lines, Ar $\mathrm{Ly\beta}$
and $\mathrm{He\beta}$, are typically larger than the core size.
Hence, the analysis of the space-resolved spectrum gives the chord-averaged
$T_{e}$ and $n_{e}$ characteristic of the sub-volume. Figure 3(b)
shows a comparison of two space-resolved spectra extracted from two
different core sub-regions. The spectra are slightly smoothed for
discussion purposes. On one hand, the spectrum from a central region
(i.e., blue) has a larger $\mathrm{Ly\beta}$ to $\mathrm{He\beta}$
ratio than that from the edge (i.e., red), which implies that chord-averaged
$T_{e}$ is higher in the central region. On the other hand, the spectrum
from the edge region shows larger broadening than that from the central
which is consistent with the idea of higher densities in the outer
region of the core.

\begin{figure}
\includegraphics[width=8.5cm]{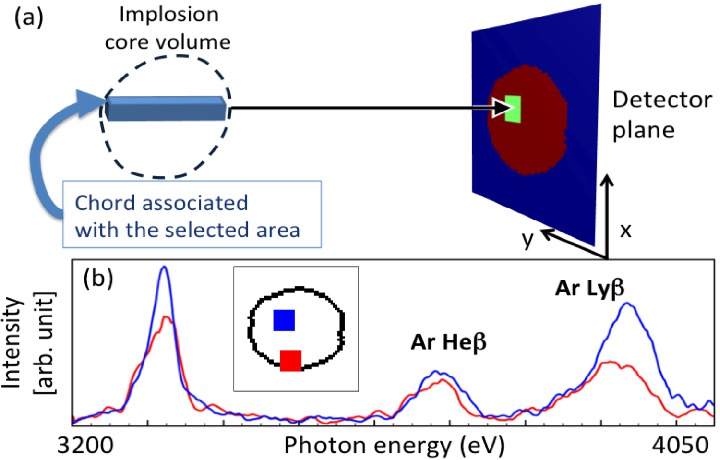} \caption{(a) Schematic to illustrate a selected image sub-region and associated
sub-volume in the implosion core. (b) Two space-resolved spectra associated
with inner (blue) and outer (red) sub-regions extracted from the data
in Fig. 2.}
\end{figure}

To model these spectra in detail, the local emissivity and opacity
of Ar were computed as a function of $T_{e}$ and $n_{e}$ using a
collisional-radiative atomic kinetics model, ABAKO \cite{2009PhRvE..80e6402F}
and detailed Stark-broadened line shapes \cite{Mancini:1991jr}. The
optical depths of He$\beta$ and Ly$\beta$ are small; still neglecting
the opacity effects in the level population kinetics and in the emergent
spectra calculation would not produce the most accurate analysis results
\cite{Keane:1993kwa,Nagayama:2008ff}. The opacity effect in the level
population kinetics is important to compute reliable plasma emissivity
and opacity. This effect was taken into account by an escape factor
approximation for a spherical source that considers the local conditions
of the plasma and the size of the core \cite{Mancini:1987wb}. The
escape factor source radius is estimated by mass conservation based
on the initial filling pressure and local density, $n_{e}$. This
approximation was tested by comparing with results from the self-consistent
solution of atomic kinetics and radiation transport equation for the
relevant plasma size and conditions \cite{Golovkin:2000ek}. The opacity
effect in the emergent spectra is taken into account by integrating
the radiation transport equation along a uniform chord, $i$:

\begin{equation}
I_{\nu}^{i}=\frac{\epsilon_{\nu}\left(T_{e}^{i},\, n_{e}^{i}\right)}{\kappa_{\nu}\left(T_{e}^{i},\, n_{e}^{i}\right)}\left[1-\exp\left(-\kappa_{\nu}\left(T_{e}^{i},\, n_{e}^{i}\right)L^{i}\right)\right]\label{eq:radiation_transport}
\end{equation}
where $\epsilon_{\nu}\left(T_{e}^{i},\, n_{e}^{i}\right)$ and $\kappa_{\nu}\left(T_{e}^{i},\, n_{e}^{i}\right)$
are the emissivity and opacity of Ar computed with the detailed atomic
model at electron temperature, $T_{e}^{i}$, and electron density,
$n_{e}^{i}$, of the chord, $i$. $L^{i}$ is the chord length estimated
by the implosion core volume extracted from the MMI data \cite{Nagayama:2011il}.
A global optimization, GALM, which consists of a genetic algorithm
(GA) followed up by a Levenberg-Marquardt non-linear least squares
minimization (LM) \cite{Nagayama:2012hh}, was used to objectively
find the optimal $T_{e}$ and $n_{e}$ of the region that produce
the best fits to the space-resolved spectrum. Since the GA is initialized
by a random number generator, running it with different initial seeds
for the same data checks the solution uniqueness. For each run, the
GA solution is fine tuned by the LM. In this way GALM combines the
complementary characteristics of the GA and the LM optimization algorithms.

To extract $T_{e}$ and $n_{e}$ spatial distributions, a total of
18 rectangular spatial regions were defined within the implosion core
boundary and the collection of space-resolved spectra associated with
each spatial region were extracted. As the sub-region size becomes
smaller, one can extract finer details about the spatial structure.
However, at the same time, the noise of the spectra becomes larger.
As one selects a smaller and smaller sub-region, at some point, the
uncertainties in the inferred conditions of the sub-region exceed
the region-to-region variations, and the extracted spatial structure
is not reliable. Thus, we selected the minimum sub-region size such
that the space-resolved spectra had enough S/N and the extracted spatial
structure is statistically meaningful. This threshold size depends
on the MMI S/N, and we found this threshold to be 16 $\mu$m for the
data discussed here. Typical S/N for Ly$\beta$ and He$\beta$ for
the extracted space-resolved spectra are 4.2 and 5.2, respectively.
With better S/N, the threshold size can become as small as the spatial
resolution of the instrument.

\begin{figure}
\includegraphics[width=8.5cm]{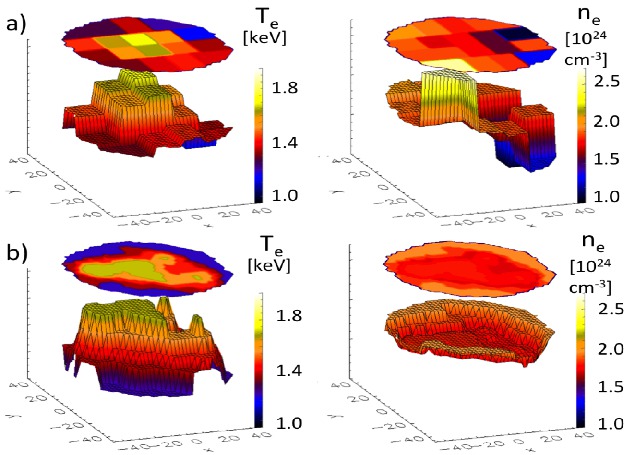}

\caption{\label{fig:results}$T_{e}$ (left) and $n_{e}$ (right) spatial distribution
results for the $t_{3}$ MMI data displayed in Fig. 2, obtained from
a collection of (a) square-type space-resolved spectra, (b) contour-type
space-resolved spectra. The x and y coordinates are given in $\mathrm{\mu m}$. }
\end{figure}

Each spectrum is independently analyzed using GALM coupled with the
spectral model described above {[}Eq. (\ref{eq:radiation_transport}){]}.
The collection of fits to the data have an average  normalized $\chi^{2}$
of 1.12 with a standard deviation of 0.35. The uncertainties were
determined by propagating the noise on the MMI data through the spectra
extraction. The analysis results for the 2-D spatial distribution
of $T_{e}$ and $n_{e}$ in the core are displayed in surface plots
and contour images in Fig. \ref{fig:results}(a). The overall shape
of the core projected onto the image plane can be approximated by
an ellipse of 91$\times$79 $\mu$m in size. Within the core, we observe
that the $T_{e}$ and $n_{e}$ distributions are counter correlated
and have asymmetric distributions. The hot spot is shifted away from
the core center by about 14 $\mathrm{\mu m}$ and has an asymmetric
structure. Also, given $T_{e}$ and $n_{e}$, we can assess how isobaric the
implosion core is based on an ideal gas approximation (i.e., $P_{e}\propto T_{e}n_{e}$).
The implosion core is isobaric within 18\%. 
The uncertainties in $T_{e}$ and $n_{e}$ inferred from
the $\chi^{2}$ analysis were $\sim3\%$ and $\sim9\%$, respectively,
and thus the extracted spatial structure is statistically significant.
We must note that these
uncertainties do not include the uncertainty in the spectral model.
Since the same model is used for all space-resolved spectra analyses,
model uncertainties would affect the results systematically without
significantly changing the relative spatial structure. We emphasize
that these $T_{e}$ and $n_{e}$ values are chord-integrated, but
they still reflect the asymmetries present in the core conditions
achieved at the collapse of the implosion.

One can envision different ways of extracting the collection of space-resolved
spectra since there is flexibility in the way in which the image sub-regions
can be defined. The ratio of the two narrow-band images, $\mathrm{Ly\beta}$
to $\mathrm{He\beta}$, is related to the temperature spatial structure
\cite{Nagayama:2008ff}. Thus, another way to define sub-region masks
is based on the contour levels of the Ar $\mathrm{Ly\beta}$ to $\mathrm{He\beta}$
ratio image. We emphasize that this ratio image is not used in the
analysis, but used only to define image sub-regions for space-resolved
spectra extraction. This alternative selection of image sub-regions
results in irregular zones but captures spatial asymmetry effectively
in fewer spatial regions with better S/N in the extracted spectra
(i.e., 7.6 and 10.0 for Ly$\beta$ and He$\beta$, respectively).
Five contour-type spatial regions are defined and the corresponding
space-resolved spectra were extracted and analyzed. The collection
of fits to the data have an average normalized $\chi^{2}$ of 1.08
with a standard deviation of 0.47. The uncertainties in the temperatures
and densities were improved to about 2\% and 5\%, respectively, due
to better S/N in the spectra. Figure \ref{fig:results}(b) shows the
resultant $T_{e}$ and $n_{e}$ distributions. This analysis helps
to study not only asymmetry in temperature but also the average densities
in the corresponding equi-temperature regions, which better illustrates
the $T_{e}$ and $n_{e}$ counter-correlation. Due to more space-averaging,
the core pressure in this representation becomes more uniform (i.e.,
within 8\%). The hot spot region in the core is approximated by an
oval of 44$\times$21 $\mu$m in size, i.e., an ellipse of larger
eccentricity than that of the core shape. Furthermore, one can more
clearly see that the hot spot is shifted away from center along the
negative x-direction and not fully surrounded by symmetric, high density
regions. 

Since this experiment was a direct-drive implosion, the combination
of laser non-uniformity, power imbalance, and target surface roughness
are all possible seeds for asymmetry. In addition, there is a stalk
that initially holds the capsule, and we noticed that its projection
on to the detector plane was along the negative x-axis in Fig. 4.
The connection between stalk (or fill tube) and core shape asymmetry
has been reported previously based on broad-band emission image data\cite{Igumenshchev:2009bm,Langer:2007fn,Hammel:2011ga,Kline:2013ea},
but our results suggest that isolated defects due to the stalk could
also correlate with the asymmetry in $T_{e}$ and $n_{e}$. Time-resolution
of the MMI data was provided by the MCP strips (Fig. 1), and the time-evolution
of the asymmetry is under investigation.

The idea of extracting space-resolved spectra from an array of spectrally
resolved images opens up a new way to analyze spatial structure of
plasmas. We presented its application to OMEGA ICF experiments and
extracted the implosion core $T_{e}$ and $n_{e}$ asymmetry for the
first time. With a suitable selection of spectroscopic tracer, the
same diagnostic method can be applied to NIF implosion experiments.
The asymmetry was investigated with two different sub-region definitions,
which provided consistent insights into the details of the spatial
structure. While the space-resolved spectra obtained from rectangular
sub-regions provides a general way to study asymmetry, the contour-type
regions help to better visualize and interpret the asymmetry information.
These two results illustrate the versatility and potential of MMI
space-resolved spectra analysis as well as the flexibility in the
sub-region selection. Finally, the extraction and analysis of space-resolved
spectra from MMI data discussed here also provides the basis to implement
a polychromatic tomography method \cite{Nagayama:2012hh} to investigate
the 3-D $T_{e}$ and $n_{e}$ distributions of laboratory high energy
density plasmas, including ICF implosion cores.

This work was supported by DOE/NLUF Grants DE-FG52-09NA29042 and DE-NA0000859,
and LLNL. RF was also partially supported by Grant No. ENE2009-11208
of the Spanish Ministry of Science and Innovation and the Keep-in-Touch
Project of the EU.

%

\end{document}